\documentclass{IEEEtran}[2015/08/26 V1.8b by Michael Shell]
\def\@IEEEclspkgerror{\ClassError{IEEEtran}}

\usepackage{cite}
\usepackage{graphicx}
\usepackage{algorithm}
\usepackage{algorithmic}
\usepackage{amsmath}
\usepackage{amssymb}
\usepackage{amsthm}

\makeatletter
\newcounter{parenttheorem}

\makeatother

\def\BibTeX{{\rm B\kern-.05em{\sc i\kern-.025em b}\kern-.08em
    T\kern-.1667em\lower.7ex\hbox{E}\kern-.125emX}}
\begin{document}
\title{Length Learning for Planar Euclidean Curves}
\author{Barak Or and Liam Hazan

\thanks{B.Or is with the Department of Marine Technologies, Charney School of Marine Sciences, University of Haifa, Israel (e-mail: barakorr@gmail.com).}
\thanks{L.Hazan is with Industrial Engineering School at Technion, Israel(e-mail:liamhaz24@gmail.com).}}

\maketitle

\begin{abstract}
 In this work, we used deep neural networks (DNNs) to solve a fundamental problem in differential geometry. One can find many closed-form expressions for calculating curvature, length, and other geometric properties in the literature. As we know these concepts, we are highly motivated to reconstruct them by using deep neural networks. In this framework, our goal is to learn geometric properties from examples. The simplest geometric object is a curve. Therefore, this work focuses on learning the length of planar sampled curves created by a sine waves dataset. For this reason, the fundamental length axioms were reconstructed using a supervised learning approach. Following these axioms a simplified DNN model, we call $ArcLengthNet$, was established. The robustness to additive noise and discretization errors were tested.
\end{abstract}

\begin{IEEEkeywords}
Differential geometry, length learning, planar Euclidean curves, supervised learning. 
\end{IEEEkeywords}

\section{Introduction}
The calculation of curve length is the major component in many classical and modern problems involving numerical differential geometry \cite{guenter1990computing,hellweg1998new}. For example, a handwritten signature involves the computation of the length along the curve \cite{ooi2016image}. Several numerical constraints affect the quality of the length calculation; additive noise, discretization error, and partial information. A robust approach to handle it is required.\\ 
Lastly, Machine Learning (ML) has become highly popular. It has achieved great success in solving many classification, regression and anomaly detection tasks \cite{lecun2015deep}. A sub-field of ML is the Deep Neural Networks (DNN), which outperforms many classic methods by design deep architectures \cite{lecun2015deep}. An efficient DNN architecture finds intrinsic properties by using a convolutional operator (and some more sophisticated operators) and generalize them. Their success is related to the enormous amount of data and their capability to optimize it by high computational available resources. In this work, we address a fundamental question in the field of differential geometry \cite{sternberg1999lectures} and we aim to reconstruct a basic property using DNN. The simplest geometric object is a curve. To characterize a curve, one can define metrics such as length and curvature, and distinguish one curve from another. There are many close form expressions for calculation of the length, curvature, and other geometric properties in the classical literature \cite{kimmel2003numerical}. However, since we know the powerful functions of DNN, we are highly motivated to reconstruct the fundamental length property for curves, the arclength, by designing a DNN. We focused on the two-dimensional Euclidean domain. The formulation of this task was done in a supervised learning method where a data-dependent learning-based approach was applied by feeding each example at a time through our DNN model and by minimizing a unique loss function that satisfies the length axioms. For simplicity, we focused on sine wave curves and created a dataset by tuning the wave amplitude, phase, translation and rotation to cover a wide-range of geometric representation. The resulted trained DNN was called $ArcLengthNet$. It obtains a 2D vector as an input, represents the planar Euclidean sampled curve, and outputs their respective length. \\ Related papers in the literature mainly address a higher level of geometric information by deep learning approach \cite{bronstein2017geometric,berg2018unified}. Saying that, a fundamental property was reconstructed by DNN in \cite{pai2016learning}, where a curvature-based invariant signature was learned by using a Siamese network configuration \cite{chicco2020siamese}. They presented the advantages of using DNN to reconstructing the curvature signature, among which it mainly results in robustness to noise and sampling errors. \\
The main contributions of this work is to reconstruct the length property. For that, two architectures were designed. One is based on Convolutional Neural Networks (CNNs), and the other is based on Recurrent Neural Networks (RNNs). We showed that the CNN-based architecture overcomes the RNN-based architecture, and by that establishing the $ArcLengthNet$ as a CNN based architecture. The advantage of the $ArcLengthNet$ is presented.\\  
The remainder of the paper is organized as follows: Section 2 summarizes the geometric background of the length properties. Section 3 provides a detailed description of the learning approach were the two architectures are presented. Section 4 presents the results followed by the discussion. Section 5 gives the conclusions.\\ 

\section{Geometric Background of Arclength}
In this section, the length properties are presented and the discretization error is reviewed.  
\subsection{Length properties}
Consider a planar parametric differential curve in the Euclidean space, $C\left( p \right) = \left\{ {x\left( p \right),y\left( p \right)} \right\}  \in {{\cal R}^2}$, where $x$ and $y$ are the curve coordinates parameterized by parameter $p \in [0,N]$, where $N$ is a partition parameter. The Euclidean length of the curve, is given by \begin{equation}
l\left( p \right) = \int_0^p {\left| {{C_{\tilde p}}\left( {\tilde p} \right)} \right|d\tilde p}  = \int_0^p {\sqrt {x_{\tilde p}^2 + y_{\tilde p}^2} d\tilde p},
\end{equation}
where  $x_p^{} = \frac{{dx}}{{dp}},y_p^{} = \frac{{dy}}{{dp}}$. Summing all the increments results in the total length of $C$, given by 
\begin{equation}
{\cal L} = \int_0^N {\left| {{C_{\tilde p}}\left( {\tilde p} \right)} \right|d\tilde p}.
\end{equation}
Following the length definition, the main length axioms are provided.\\
{\bf Additivity}: The length additives with respect to concatenation, where for any  $C_1$ and $C_2$ the following holds 
\begin{equation}
{\cal L}\left( {{C_1}} \right) + {\cal L}\left( {{C_2}} \right) = {\cal L}\left( {{C_1} \cup {C_2}} \right).
\end{equation}
{\bf Invariant}: length is invariant  with respect to rotation ($\bf R$) and translation ($\bf T$), 
\begin{equation}
{\cal L}\left( {\left( {{\bf{T}} + {\bf{R}}} \right)C} \right) = {\cal L}\left( C \right).
\end{equation}
{\bf Monotonic}: length is monotone, where for any $C_1$ and $C_2$ the following holds
\begin{equation}
{\cal L}\left( {{C_1}} \right) \le {\cal L}\left( {{C_2}} \right)\,\,\,\,\,\,\,\,{C_1} \subset {C_2}.
\end{equation}
{\bf Non-negativity}: The length of any curve is non-negative, 
\begin{equation}
{\cal L}\left( C \right) \ge 0.  
\end{equation}
In order to reconstruct the length property by DNN, a discretization of the curve should be applied. As a consequence, it prone to errors.
\subsection{Discretization error}
The curve $C$ lies on a close interval $[\alpha,\beta]$. In order to find the length by a discretized process, a partition of the interval is done, where
\begin{equation}
{\cal P} = \left\{ {\alpha  = {p_0} < {p_1} < {p_2} <  \cdots  < {p_N} = \beta } \right\}.
\end{equation}
For every partition ${\cal P}$, the curve length can be represented by the sum
\begin{equation}
s\left( {{\cal P}} \right) = \sum\limits_{n = 1}^N {\left| {C\left( {{p_n}} \right) - C\left( {{p_{n - 1}}} \right)} \right|}.
\end{equation}
The discretization error is given by, 
\begin{equation}
\begin{array}{l}
{e_d} = {\cal L} - s\left( {\cal P} \right)\\
 = \int_0^N {\left| {{C_p}\left( p \right)} \right|dp - } \sum\limits_{n = 1}^N {\left| {C\left( {{p_n}} \right) - C\left( {{p_{n - 1}}} \right)} \right|} 
\end{array}.
\end{equation}
where obviously, $e_d \to 0$ when $N \to \infty$ (for further reading, the reader refers to \cite{do2016differential}). Fig. 1 illustrates a general curve with their discretized representation.

\begin{figure}[!t]
\centering
\includegraphics[scale=0.18]{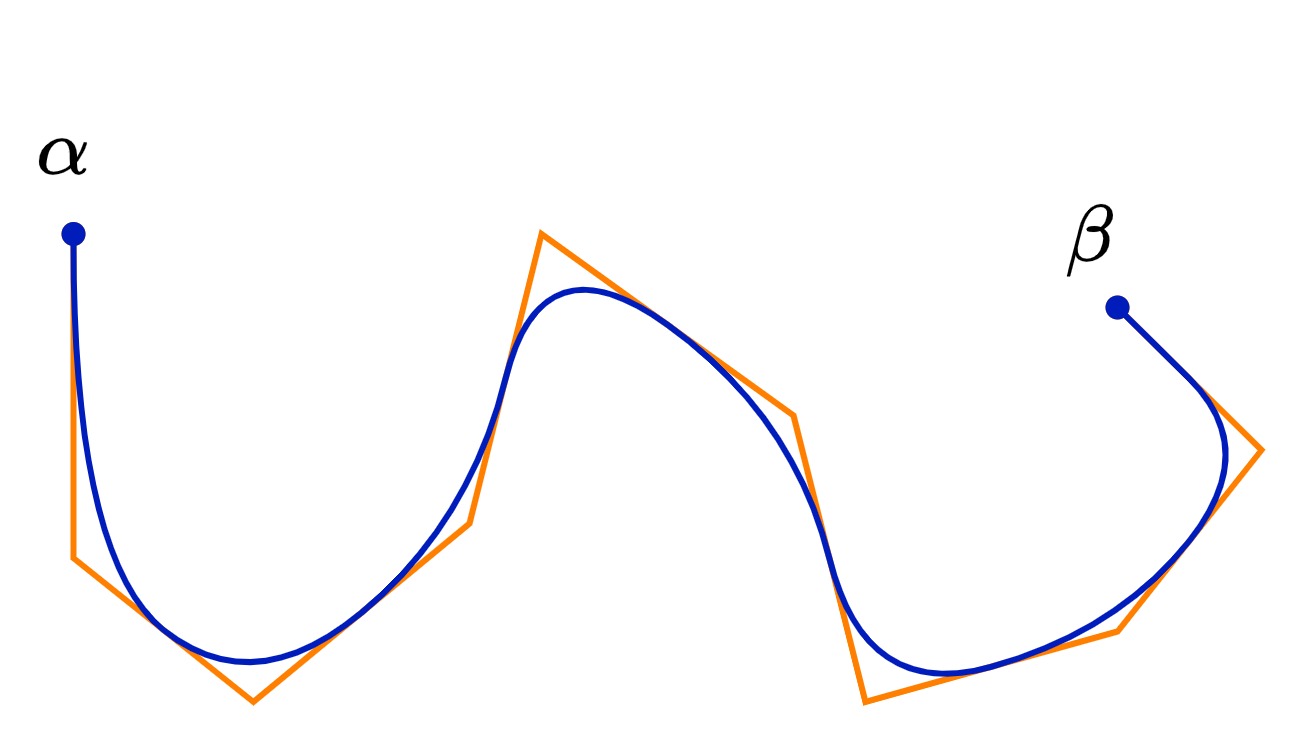}
\caption{Discretization}
\end{figure}

\section{Learning approach}
\subsection{Motivation} The motivation for using DNN for this task lies in the core challenge of implementing equations (1) and (2) in real-life scenarios. These equations involve nonlinearity and derivatives. Poor sampling and additive noise might lead to numerical errors \cite{qian1993wavelets}. The differential and integral operators can be obtained by using convolution filters \cite{pai2016learning}. The differential invariants can be interpreted as a high pass filter and the integration as a low pass filter. Hence, it is convenient to use the Convolutional Neural Network (CNN) for our task. Another approach to deal with this task involves the Recurrent Neural Network (RNN), where the curve is considered a time-series \cite{tsoi1997discrete,smagulova2019survey}. A modified version of RNN is the Long-Short Term Memory (LSTM), where a weighting for the past time steps is employed \cite{hochreiter1997long}. We implemented two architectures. The first is based on a simplified CNN, and the second is based on a simplified LSTM. We found that the CNN-based architecture overcomes the LSTM-based architecture. The rest of this section presents the details of our process to obtained both. 
\subsection{Data generation} 
The reconstruction of the length properties was done in a supervised learning approach, where many curve examples with their lengths as labels were considered. Each curve is represented by $2 \times N$ vector for the $x$ and $y$ coordinates and a fixed number of points $N$. \\
We created a dataset with $20,000$ to fully enable DNN training. This large amount of examples aimed to cover curve transformations and to satisfy different patterns. These curves were created by taking a sinus wave with a random sampling of amplitude $a$, phase $\phi$, translation $\bf T$, and rotation, $\bf R$. The general curve in our data is given by
\begin{equation}
C\left( p \right) = {\bf{R}}\tilde C\left( p \right) + {\bf{T}},
\end{equation}
where
\begin{equation}
\tilde C\left( p \right) = a\sin \left( {p + \phi } \right).
\end{equation}
A random cutting point was randomly selected to divide the curve into two new curves to reconstruct the additivity property. Fig. 2 shows some curves we created. 
The data set was split into train and test sets, where additional holdout set was created. During the data creation phase, we demanded non-negativity of the labels and created many curves of different length with rotation and translation examples to cover the various axioms (3)-(6). \\
\subsection{Loss function}
In order to show the additive property, we designed a unique loss function for each example, where it designed as follows
\begin{equation}
{J_k} = {\left( {{\cal L}\left( {{s_1}} \right) - O\left( {{s_2}} \right) - O\left( {{s_3}} \right)} \right)^2}_k + \lambda \left\| {{\Theta _{ij}}} \right\|_2^2,
\end{equation}
where ${s_1},{s_2}$, and ${s_3}$ are the input curves that hold the equality ${\cal L}\left( {{s_1}} \right) = {\cal L}\left( {{s_2}} \right) + {\cal L}\left( {{s_3}} \right)$, $O$ is the DNN output, $k$ is the example index, $\lambda$ is a regularization parameter, $\Theta_{ij}$ are the various DNN weight, and $\left\|  \cdot  \right\|_2^2$ is the two dimension norm. By minimizing $J_k$ by passing the examples through a model, the weights were tuned. The optimized model is characterized by the optimal weights that provided by
\begin{equation}
\Theta _{ij}^* = \mathop {\arg \min }\limits_{{\Theta _{ij}}} \sum\limits_k {{J_k}}.
\end{equation}

\begin{figure}[!t]
\centering
\includegraphics[scale=0.58]{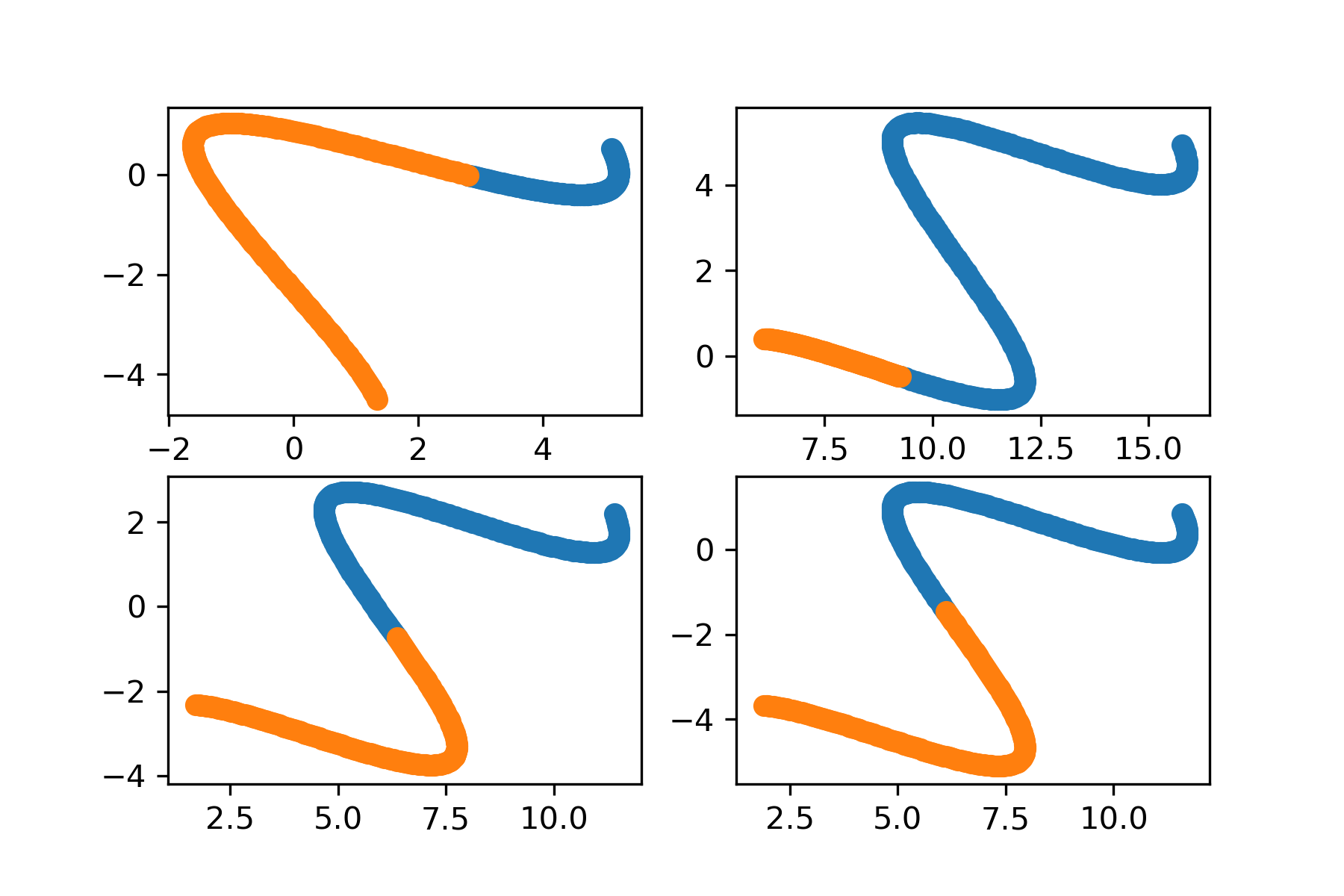}
\caption{Curve examples}
\end{figure}

\subsection{ArcLengthNet Architecture}
In this subsection the $ArcLengthNet$ is presented (the CNN-based architecture). The model architecture is very simplified, including a convolutional layer and two fully-connected layers with only one activation function. Each curve is represented by $N=200$ points. This representation is inserted into a convolutional layer with a small kernel of size $3$. It is processed into a fully connected layer that outputs only $10$ weights through a Rectified Linear Unit (ReLU) activation function to another fully connected layer which finally outputs the length. The architecture is shown in Fig.3.

\begin{figure}[!t]
\centering
\includegraphics[scale=.17]{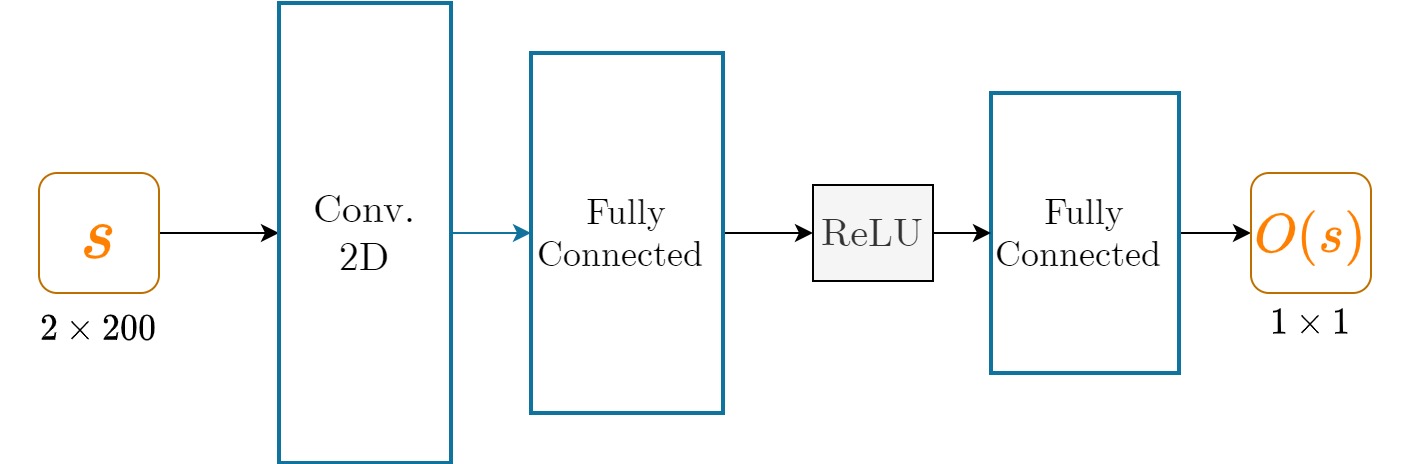}
\caption{ArcLengthNet architecture}
\end{figure}

\begin{figure}[!t]
\centering
\includegraphics[scale=0.55]{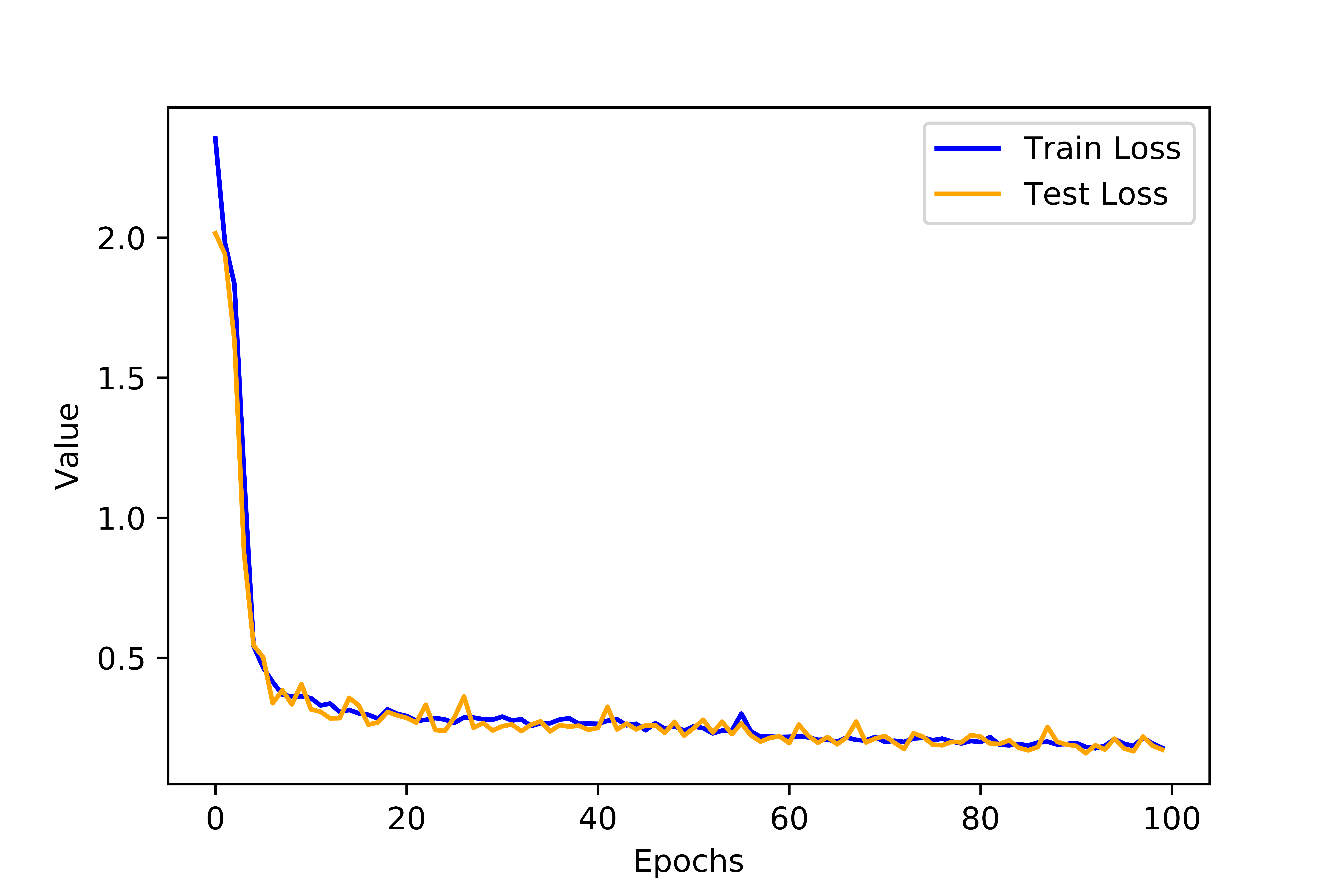}
\caption{ArcLengthNet training}
\end{figure}

\subsection{$ArcLengthNet$ Training}
The DNN was trained by passing many examples in small batches with a back-propagation method. The training process was carried out in batches of $200$ examples for $100$ epochs. The optimizer we used is Stochastic Gradient Descent (SGD) with momentum and weight decay \cite{ruder2016overview}. Various parameters are provided in Table 1. Fig.4 shows a graph of the train and test losses as a function of the number of epochs. 

\subsection{LSTM-based Architecture}
The task of length learning can be interpreted as a time-series based learning. 
Each point of the curve can be interpreted as a time step with $x$ and $y$ coordinates. The DNN aims to generalize the local properties into a global length. The typical architecture to deal with time-series data is the Recurrent Neural Network (RNN). One modification of RNN is the Long Short-Term Memory (LSTM) architecture, where has feedback connections (in opposite to the classical RNN). The LSTM is used in handwritten recognition tasks with a great success \cite{graves2008novel}.
Motivated by this success, we designed an LSTM architecture, presented in Fig.5.  Similar to $ArcLengthNet$, the LSTM-based  architecture is simplified, including $200$ blocks with  $1\time4$ vector outputs. They concatenated and inserted to two fully connected layers ($10 \time 1$ each) that outputs connected with a ReLU activation function. Fig.6 shows a graph of train and test losses as a function of the number of epochs. 

\begin{figure}[!t]
\centering
\includegraphics[scale=0.14]{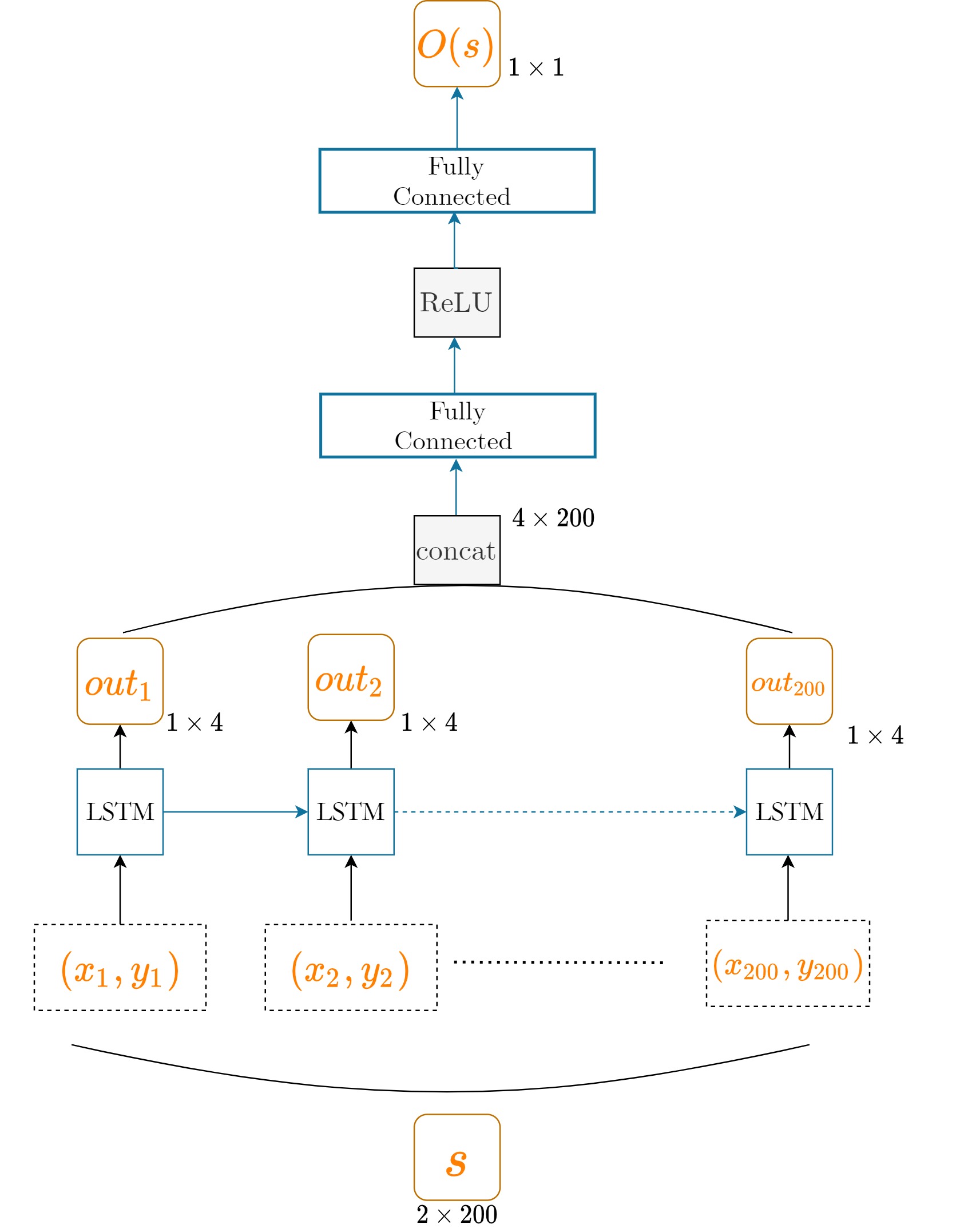} 
\caption{LSTM-based architecture}
\end{figure}

\begin{figure}[!t]
\centering
\includegraphics[scale=0.58]{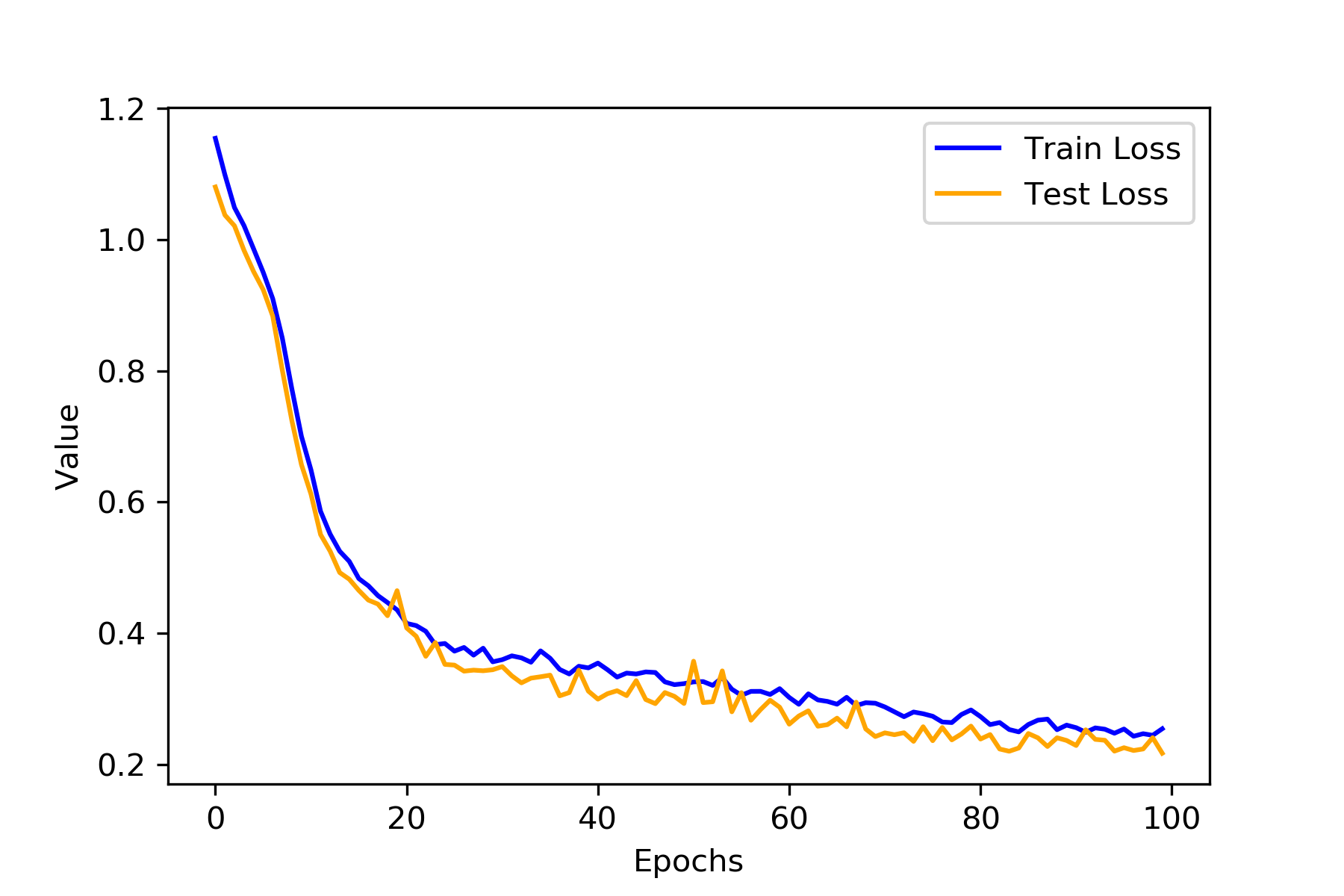} 
\caption{LSTM-based architecture training}
\end{figure}

\begin{table}[ht!]
\caption {Learning  Parameters} \label{tab:title} 

\begin{center}
\begin{tabular}{ |c|c|c|c| } 
\hline
Description & Symbol & Value \\
\hline
Nunber of examples       & $K$      & $20,000$   \\ 
Train/test ratio                   &      -     & $80/20$      \\ 
Regularization parameter                  & $\lambda$        & $0.01$ \\ 
Partition parameter      & $N$         & $200$      \\ 
Batch size            & -        & $200$      \\ 
Learning rate   & $\eta$           & $0.001$ \\ 
Momentum         & $-$           & $0.9$              \\ 
Weight decay  & -  & $0.0005$          \\ 
Epochs & -  & $100$         \\ 
\hline
\end{tabular}
\end{center}
\end{table}

\section{Results and discussion}
Both architectures were trained in the same approach with the same database. As shown in Fig.3 and Fig.6 these architectures were well trained after $100$ epochs. A holdout set was defined to test the  performance of the architectures on unseen data. This set contains $5,000$ examples that have not been used in  train set or test test. The $ArcLengthNet$ obtained a minimum  MSE of $0.17$, where the LSTM based model obtained a minimum MSE of $0.24$. Hence, we conclude that the $ArclengthNet$ architecture overcomes LSTM-based architecture. The monotonic property was tested for the $ArcLengthNet$ on this holdout set, where a linear relation was established between the true length and the $ArcLengthNet$ (Fig.4).  

\begin{figure}[!t]
\centering
\includegraphics[scale=0.58]{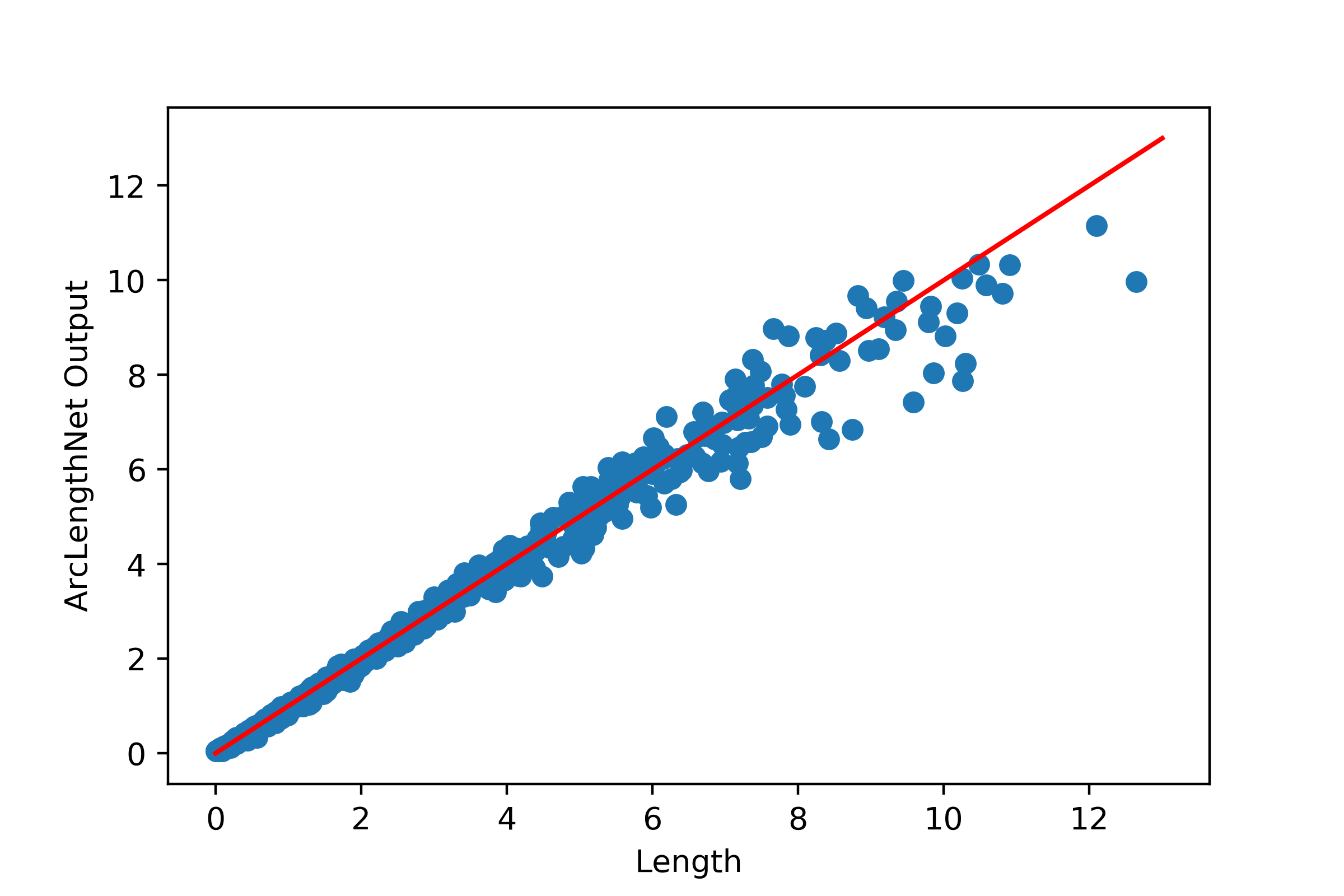} 
\caption{Monotonic property on holdout set ($ArcLengthNet$)}
\end{figure}

\subsection{Performance measure}
In order to verify our results, we used the Mean Squared Error (MSE) criterion, Root-MSE (RMSE), as also a unique criterion: RMSE-over-Mean-Length-Ratio (RMLR), given by
\begin{equation}
{\bf{RMLR}} = \frac{{{\bf{RMSE}}}}{{{\cal E}\left[ {\cal L} \right]}},
\end{equation}
where ${\cal E}$ is the expected value operator. This measure provides a normalized error with respect to the curve length.various curves of different lengths, we must appropriately weigh the errors.

\section{Conclusion and further work}
A learning-based approach to reconstruct the length of curves was presented. The powerful of deep neural networks to reconstruct the fundamental axioms was demonstrated. The results can be further used to improve handwritten signatures, and reconstruct some more differential geometry properties and theorems.

\bibliographystyle{IEEEtran}
\bibliography{all}

\end{document}